\documentclass[10pt,letterpaper,twocolumn]{article}
\usepackage{ol2}
\usepackage{graphicx}
\usepackage[T1]{fontenc}
\usepackage{subfigure}
\usepackage{amsmath}
\usepackage{amsfonts}
\usepackage{stmaryrd}
\usepackage{psfrag}
\usepackage{tabularx}\newcolumntype{Y}{>{\centering\arraybackslash}X}
\graphicspath{{/home/image/eps/pmma/}}
\DeclareGraphicsExtensions{.eps,.EPS,.ps,.PS,.eps.gz}
\begin{document}
\twocolumn[
\title{Far-field mapping of the longitudinal magnetic and electric optical fields}

\author{C. Ecoffey, T. Grosjean}

\affiliation{Département d'Optique P.M. Duffieux,\\ Institut
FEMTO-ST, UMR CNRS 6174, Universit\'e  de Franche-Comt\'e,\\ 16
route de Gray, 25030 Besançon cedex, France}

\begin{abstract}

In this letter, we demonstrate the experimental mapping of the longitudinal magnetic and electric optical fields with a standard scanning microscope
that involves a high numerical aperture far-field objective. The imaging concept relies upon the insertion of an azimuthal or a radial polarizer
within the detection path of the microscope which acts as an optical electromagnetic filter aimed at transmitting selectively to the detector the signal from the magnetic or electric longitudinal fields present in the detection volume, respectively. The resulting system is thus versatile, non invasive, of high resolution, and shows high detection efficiencies. Magnetic optical properties of physical and biological micro and nano-structures may thus be revealed with a far-field microscope.\\

\end{abstract}

\maketitle ]


In paraxial regime, light distributions are generally considered as purely transverse fields and are often described with scalar theory.  When leaving paraxial conditions for approaching the wavelength scale, the longitudinal components of the electric and magnetic fields (parallel to the propagation direction) become noticeable and light has to be seen as a 3D vectorial electromagnetic field. The enhanced longitudinal electric field produced by a radially polarized beam in a focal region has found a large interest in various domains such as particle acceleration \cite{fontana:jap83}, laser cutting \cite{niziev:jpd99}, lithography \cite{grosjean:ol07}, far-field \cite{quabis:oc00,novotny:prl01} and near-field \cite{bouhelier:prl03,stockle:cpl00,bharadwaj:aop09} microscopies and spectroscopies.  Azimuthal polarization show the inverse electromagnetic configuration for which focal spots show enhanced longitudinal optical magnetic fields \cite{youngworth:ox00}. Such beams have been successfully used to excite the magnetic resonances of split ring resonators \cite{banzer:ox10} or generate purely longitudinally polarized needles of optical magnetic field \cite{grosjean:oc13}.

Rigorous numerical methods, such as FDTD (Finite Difference Time Domain) \cite{taflove:book} or plane wave spectrum \cite{bornwolf}, predict strong discrepancies between the spatial distributions of longitudinal and transverse fields.  Various techniques have been then proposed to probe selectively the longitudinal electric and magnetic optical fields generated in a focal spot or right at the surface of an optical structure, such as single molecules \cite{novotny:prl01}, azo-dye polymers \cite{grosjean:ox06} and near-field probes \cite{hayazawa:apl04,bouhelier:prl03,lee:natphot07,burresi:science09}. Longitudinal light fields have also been deduced from the measurement of the amplitude and phase of the transverse electric field \cite{grosjean:ox10}. All these techniques allow for mapping subwavelength distributions of the longitudinal electric or magnetic fields but require the development of non trivial optical probes and/or acquisition procedures.

In this paper, we demonstrate the selective mapping of the longitudinal electric and magnetic optical fields with a standard far-field scanning microscope configuration that involves a conventional microscope objective. These selective electromagnetic detections are achieved by inserting a radial or an azimuthal polarizer within the detection path of the microscope, respectively. The resulting system is thus versatile, non invasive, and shows high detection efficiencies. 
Such a vectorial far-field imaging should enable detailed electromagnetic information onto highly confined light field distributions, impacting applications such as biosensing, optical trapping, metamaterial analysis, focusing and beam shaping.



Under focusing, radially and azimuthally polarized beams are known to produce single confinements of longitudinal electric and magnetic fields at their center, respectively \cite{youngworth:ox00}. These confined longitudinal fields are surrounded by transverse field distributions which can be deduced from the longitudinal fields with Maxwell's equations \cite{bornwolf}. According to Maxwell's Ampere equation (Maxwell's Faraday equation, respectively), a single confinement of longitudinal electric field (longitudinal magnetic field, respectively) generates a loop of transverse magnetic field (transverse electric field, respectively), and vice versa. In linear polarization, focal spots show a single confinement of linearly-polarized transverse electric field at their center. From Maxwell's equations, the rest of the electromagnetic field in the focal region can be seen as being induced by this single transverse field confinement.

Following reciprocity, the signal from the longitudinal component of the electric and magnetic optical fields present at the center of the detection volume of an objective used in collection mode are transmitted through the objective as radially and azimuthally polarized beams, respectively, or fields distributions which overlap these axially polarized beams. As a comparison, the signal from the transverse electric field is transmitted by an objective as a linearly polarized beam. We propose here to map selectively the longitudinal electric and magnetic optical fields scattered by physical and biological samples by extracting selectively from the signal collected by an objective the information that propagates toward the detector under the form of radially and azimuthally polarized beams, respectively, or beams with similar axial polarizations.

\begin{figure}[htbp]
\begin{center}
\includegraphics [width=0.99\columnwidth]{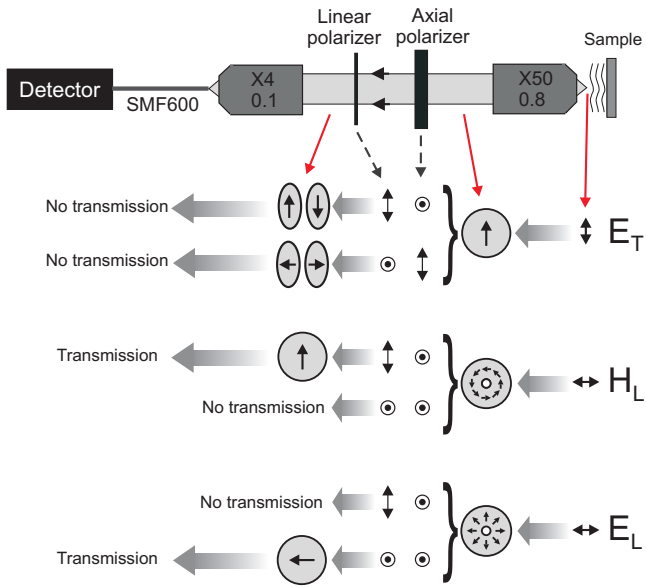}
\caption{Scheme of the far-field detection system and principle of the detection process.}\label{fig:schema}
\end{center}
\end{figure}

To this end, we developed the far-field detection systems shown in Fig. \ref{fig:schema}. It combines the detection bench of a confocal microscope to  an axial (radial or azimuthal) and a linear polarizer. The light fields collected locally by a high numerical aperture (NA) objective ($\times$50, 0.8) is projected onto a second objective of lower NA ($\times$4, 0.1) to be efficiently injected into a monomode fiber connected to a photodetector. The entrance facet of the optical fiber is conjugated optically to the focus of the high NA objective: the detection volume of the system coincides with the focal region of the high-NA objective. Radial and azimuthal polarizers are usually used to convert a linearly polarized light beam into radially and azimuthally polarized doughnut beams, respectively. Following reciprocity, we propose here to use these polarizers (inserted in between the two objectives) in order to convert radially and azimuthally polarized beams into linearly polarized beams.  The linear polarizer, placed after the axial polarizer in the detection path, is used as an analyzer to transmit or stop the signal from these incident axially polarized incident beams.
%

The axial polarizer is a passive and compact component used to convert an incident linearly polarized beam into either a radially polarized or an azimuthally polarized beam. When the incident polarization is parallel to the axial polarizer's axis, a radially polarized beam is generated. A perpendicular incident polarization direction leads to an azimuthally polarized doughnut beam. From reciprocity, this axial polarizer coupled to a linear polarizer transmits power from incident radially or azimuthally polarized beams if the two polarizers' axis are parallel or perpendicular, respectively.

The detection principle of our system is schemed in Fig. \ref{fig:schema}. The transverse components of the electric field ($E_T$) present at the center of the detection volume are converted into linearly-polarized collimated beams by the first objective. Depending on the orientation of the axial polarizer with respect to the incident polarization direction, a radially or an azimuthally polarized beam is generated, which is converted into a linearly polarized hermite-gauss beam by the linear polarizer. Such beams, which couple into the $TE_{01}$, $TM_{01}$ and $HE_{21}$ fiber modes \cite{grosjean:oc05}, cannot be transmitted through the single-mode fiber piece connected to the detector (which transmits only the fundamental mode $HE_{11}$). Therefore, the signal from the transverse electric field cannot be detected. As noted previously, the longitudinal electric field (magnetic field) at the center of the detection volume of our system leaves the first objective as a radially (azimuthally) polarized beam. Such a beam is converted by the axial polarizer into a linearly polarized beam along (perpendicular to) the axial polarizer's axis. If the linear and axial polarizers have parallel (perpendicular) axis, the signal is transmitted to the fiber, couples to its fundamental $HE_{11}$ mode, and a signal is detected. If the axis are perpendicular (parallel), the linear polarizer stops the signal. Therefore, the imaging system proposed here can detect either the longitudinal electric or magnetic field simply by rotating by 90$^{\circ}$ the linear polarizer with respect to the axial polarizer's axis, or vice versa.

In the following, a He-Ne  collimated laser beam ($\lambda=632.8$ nm) is used to illuminate the sample from the backside  and the transmitted optical field distribution is mapped while the sample is raster scanned with a 3D piezo stage. The photodetector (a photomultiplier tube), connected to a computer, records the optical signal transmitted by the detection bench at each raster point, leading to 2D transverse or longitudinal maps of optical fields.

\begin{figure}[h!]
\begin{center}
\includegraphics [width=0.99\columnwidth]{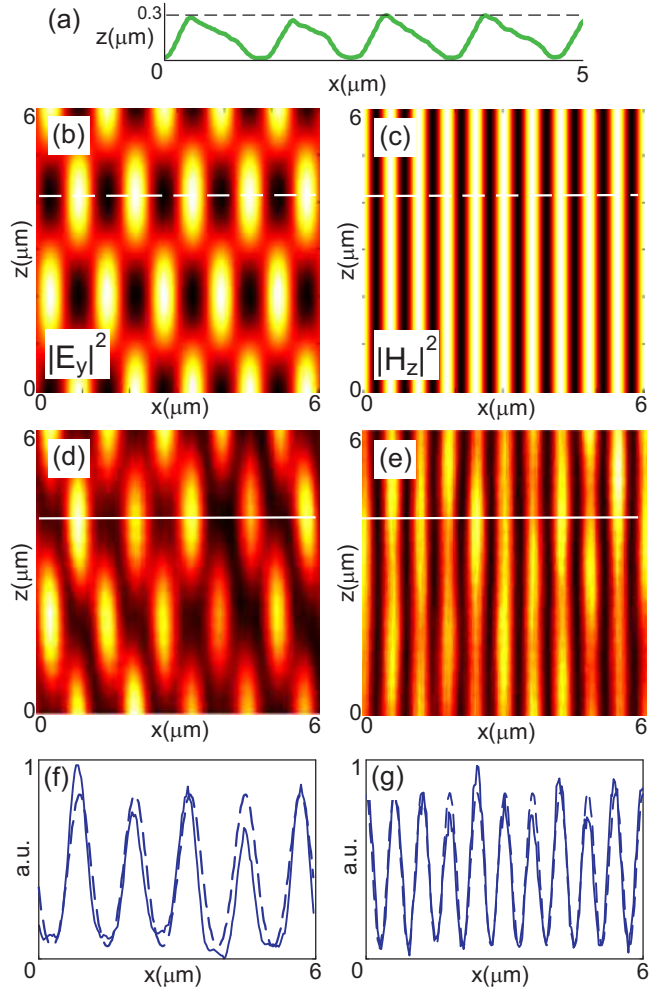}
\caption{(a) Grating's profile along (xz)-plane measured by atomic force microscopy. This dielectric sample is used in transmission mode and contains 830 grooves per mm aligned along (0y)-axis. (b) and (c) Simulation in a longitudinal (xz)-plane of the intensity distribution of (b) $E_y$ and (c) $H_z$ produced by the grating at $\lambda=632.8$ nm and with a s-polarized incident plane wave. (d) and (e) Experimental acquisitions of the grating diffraction pattern in a longitudinal (xz)-plane (d) without and (e) with azimuthal polarizer in the detection path of the microscope (the linear polarizer is oriented along (0y)). The grating is illuminated from the backside  with a s-polarized He-Ne laser beam at normal incidence. (f) and (g) Theoretical (dashed curves) and experimental (solid curves) profiles of the grating's diffracted field, (f) along the lines shown in (b) and (d) (electric field), and (g) along the lines shown in (c) and (e) (magnetic field).}\label{fig:grating}
\end{center}
\end{figure}

First, our far-field imaging concept is tested onto a 1D dielectric surface grating whose profile, measured by atomic force microscopy, is shown in Fig. \ref{fig:grating}(a). The grating, whose invariance direction is along (0y), is illuminated from the backside with a collimated s-polarized He-Ne laser beam at normal incidence (incident waves propagate along (0z)). Under these conditions, the grating generates 3 homogeneous diffraction orders in air which produce an interferogram that involves only 3 vectorial components of the optical field $E_y$, $H_x$ and $H_z$. Figs \ref{fig:grating}(b) and (c) show the simulation of the intensity distributions of $E_y$ and $H_z$ diffracted by the sample. These two field distributions can be easily distinguished from each other and can thus be used as test-field distributions to validate our microscopy concept. 
Figs. \ref{fig:grating}(d) and (e) display the experimental acquisition mappings of the field diffracted by the grating (d) with and (e) without the azimuthal polarizer in the detection path of the microscope, the linear polarizer being oriented along (0y). These two images have been acquired successively over the same area along the longitudinal plane (xz). 
We see that the two experimental images are in good agreement with the simulations of $|E_y|^2$ and $|H_z|^2$ shown in Figs. \ref{fig:grating}(b) and (c), respectively. This agreement is confirmed in Figs. \ref{fig:grating}(f) and (g) which compare theoretical profiles of $|E_y|^2$ and $|H_z|^2$ (along the dashed lines of Figs. \ref{fig:grating}(b) and (c)) to the profiles of the experimental images acquired without and with the azimuthal polarizer (along the solid lines of Figs. \ref{fig:grating}(d) and (e)). This unambiguously prove that our far-field microscope is capable of mapping the magnetic field of light ($H_z$) diffracted by 1D samples.
%

\begin{figure}[h!]
\centering
\includegraphics[width=0.99\columnwidth]{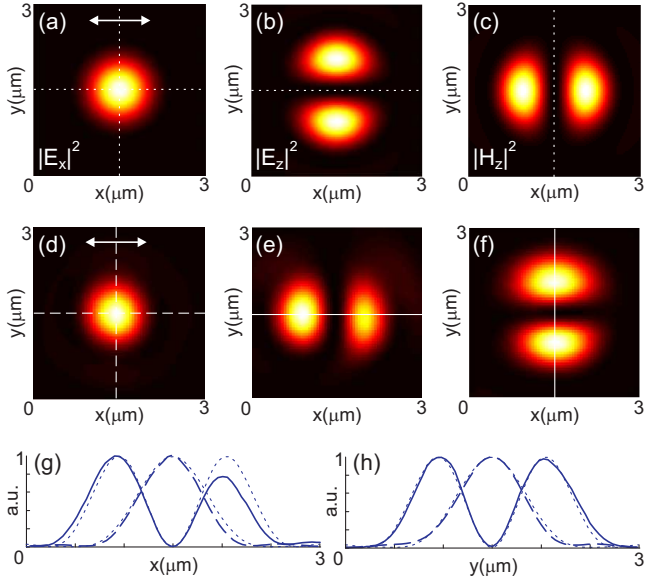}
\caption{(a-c) Simulation of the intensity distributions of (a) $E_x$, (b) $E_z$ and (c) $H_z$ at the focus of an objective (NA=0.45) illuminated with an incident beam linearly polarized along (0x) ((0z) is the longitudinal axis). The intensity distributions are plotted along the transverse (xy)-plane at $\lambda=632.8$ nm. (d-f) Experimental acquisitions of the diffraction pattern produced at the focal plane of a ($\times$20, NA=0.45) objective by an incident HeNe laser beam, linearly polarized along (0x), (d) with a single linear polarizer oriented along (0x), (e,f) with the combinations of linear/radial and linear/azimuthal polarizers, respectively. (g) profiles along (0x) of (d) (dashed curve), (e) (solid curve) and (a,b) ($|E_x|^2$ and $|E_z|^2$, dotted curves). (h) profiles along (0y) of (d) (dashed curve), (f) (solid curve) and (a,c) ($|E_x|^2$ and $|H_z|^2$, dotted curves).}
\label{fig:focus}		
\end{figure}

In a second time, our system is used to probe the light field at the focus of an objective. Focused beams offer the unique opportunity to produce simple and reproducible 2D field distributions in a transverse plane that can be easily identified and predicted theoretically. Figures \ref{fig:focus}(a-c) show the simulation of the transverse field distribution generated at the focus of an objective of NA=0.45 by an incident collimated beam that is linearly-polarized along (0x) \cite{Richards:59}. The field amplitude at the objective's pupil plane is described by a gaussian function whose $1/e$ width matches the pupil diameter. The intensity distribution of $E_x$ leads to a single spot (Fig. \ref{fig:focus}(a)) whereas $E_z$ and $H_z$  (longitudinal fields) generate orthogonal two-spots structures, as shown in Figs. \ref{fig:focus}(b) and (c). Therefore, linearly polarized focused beams allow for validating our imaging concept to the mapping of 2D optical field distributions in a transverse plane.

Figures \ref{fig:focus}(d-f) report the experimental mapping of the focal spot generated by a ($\times 20,0.45$) objective illuminated with a linearly polarized HeNe laser beam along (0x). The laser beam is launched toward the objective with an optical fiber coupled to a collimation lens and the overall optical bench is raster scanned in the transverse (xy)-plane with a piezo stage during image acquisition. Here again, when only a single linear polarizer is inserted in the microscope detection path, the experimental image (Fig. \ref{fig:focus}(d)) is in good agreement with the simulation of $|E_x|^2$. When the radial and azimuthal polarizers are added, the experimental acquisitions (Figs. \ref{fig:focus}(e) and (f)) reveal $|E_z|^2$ and $|H_z|^2$, respectively (cf. Figs. \ref{fig:focus}(b) and (c)). The good agreement between the experimental mappings and theoretical predictions of the fields at focus is confirmed in Figs. \ref{fig:focus}(g,h) which plot the experimental and theoretical intensity profiles across the focal region, along the solid, dashed and dotted lines shown in Figs. \ref{fig:focus}(a-f).


The results shown in this paper demonstrate the ability of our far-field microscope to locally map 3D distributions of longitudinal magnetic and electric light fields. In other words, we show that a sample can be imaged optically through the selective detection in the far-field  of either the longitudinal electric or magnetic fields. It may thus be possible to reveal for example the magnetic optical properties of physical and biological samples with a far-field microscope. This detection concept is independent of the illumination process of the sample, and can therefore be used in a standard scanning far-field microscope to map large scale light field distributions, or it can be integrated in a confocal microscope \cite{wilson:ap90}. New polarization contrast confocal microscopy could then be imagined, involving longitudinal electric and magnetic fields in the detection. It is also possible to realize the simultaneous mappings of the longitudinal and transverse optical electric fields and longitudinal magnetic field by implementing a multichannel acquisition system. The resolution of the microscope, about $0.75 \lambda$ in our case, can be improved up to $0.4 \lambda$ by using immersion objectives. Note that the image forming process shown here may require vector diffraction theories to be analyzed, instead of scalar diffraction theories as usually employed in far-field microscopies \cite{bornwolf,goodman}. An analytical description of the microscope with a vectorial transfer function may be possible and would merit further investigations. Finally, the selective coupling to the detector of the information that is collected under the form of waves showing hybrid polarizations \cite{grosjean:oc05} would allow for accessing the quadrupolar optical information of a sample. This information, which is often hidden by the dipolar information, would provide new insight onto light-matter interaction.

The authors are indebted to Remo Giust for helpful discussions. This work is supported by the
"Pôle de compétitivité Microtechnique" and the Labex ACTION.


\end{document}